\documentclass[10pt,a4paper]{article}

\usepackage[top=2cm, bottom=2cm, left=1.4cm, right=1.7cm, paperwidth=16.2cm, paperheight=24.2cm]{geometry}
\usepackage{amsmath}
\usepackage{amsthm}
\usepackage{fancybox}
\usepackage{bbm}
\usepackage{pdflscape}
\usepackage{algorithm}
\usepackage{algpseudocode}
\usepackage{caption}
\usepackage{subcaption}
\usepackage{placeins}
\usepackage{array}
\usepackage{multirow}
\usepackage{marvosym}
\usepackage{longtable}
\usepackage[version=3]{mhchem}
\usepackage{natbib}
\usepackage{textcomp}
\usepackage{rotating}
\usepackage{enumitem}
\usepackage{pifont}
\usepackage{url}

\newcommand{\E}{\mathbbm{E}}

\newcommand{\F} {\ensuremath{\mathcal{F}}{}\!}

\newcommand{\h}{\ensuremath{\frac{1}{2}}}
\newcommand{\Vhat}{\ensuremath{\widehat{V}}}

\begin{document}
\author{Andrew Green\footnote{Contact: andrew.green2@scotiabank.com. Andrew Green is a Managing Director and lead XVA Quant at Scotiabank\texttrademark\, in London. The views expressed in this article are the personal views of the author and do not necessarily reflect those of Scotiabank\texttrademark. \texttrademark\, Trademark of The Bank of Nova Scotia and used under license. Important legal information and additional information on the trademark may be accessed here: http://www.gbm.scotiabank.com/LegalNotices.htm} and Chris Kenyon\footnote{Contact: chris.kenyon@lloydsbanking.com The views expressed are those of the author only, no other representation should be attributed.   Not guaranteed fit for any purpose.  Use at your own risk.}}
\title{XVA at the Exercise Boundary}
\date{24 September 2016\vskip5mm Version 0.8}

\maketitle

\begin{abstract}
XVA is a material component of a trade valuation and hence it must impact the decision to exercise options within a given netting set. This is true for both unsecured trades and secured / cleared trades where KVA and MVA play a material role even if CVA and FVA do not. However, this effect has frequently been ignored in XVA models and indeed in exercise decisions made by option owners. This paper describes how XVA impacts the exercise decision and how this can be readily evaluated using regression techniques \citep{longstaff2001valuing}. The paper then assesses the materiality of the impact of XVA at the exercise boundary on swaption examples.
\end{abstract}

\section{Introduction}

XVAs have become an increasing important part of derivative valuation \citep[See for example][and references therein]{kenyon2016landmarks}. Hence XVA should also be part of option exercise decisions except where cash settlement takes place immediately after exercise of the option.\footnote{For example a cash settled European swaption settles with a cash payment between 0 and 2 days after exercise. The XVAs associated with such a short period will likely not be material.} The no exercise value, or continuation value, for Bermudan and American style options will contain the XVAs associated with not exercising the option, while the exercise value will contain the XVAs associated with exercising the option. The presence of XVAs will, therefore, clearly shift the exercise boundary of the option. This will be potentially significant for trades with long maturity underlying trades such as physically settled European swaptions. Hence the basic option exercise criterion that
\begin{equation}\label{e:ex1}
V_{ex}(t) > V_{noex}(t)											
\end{equation}
where $V_{ex}$ is the XVA-free exercise value and $V_{noex}$ is the XVA-free no-exercise value is replaced by
\begin{align}\label{e:ex2}
\Vhat_{ex}(t) > & \Vhat_{noex}(t)\\	
V_{ex}(t) + U_{ex}(t) >& V_{noex}(t) + U_{noex}(t)
\end{align}
where $\Vhat$ is the economic value and $U$ is the valuation adjustment. 

Given that change in exercise boundary due to XVAs, then the option value itself will also change. Hence XVAs impact both option decision and option valuation. As the valuation is impacted then the implied option smile will also change.

Furthermore valuing the trade at some time $t < t_{ex}$ will require a model than can estimate the shift of the exercise boundary. This shift is determined by future XVAs, which are themselves expectations and hence it should be clear that a forward conditional expectation calculation will be required. This is of course computationally challenging and normally approached using regression techniques. 

Unfortunately, in general, XVA adjustments are calculated at units greater than the individual trade. For example, CVA / DVA and FVA are calculated at netting set level, while KVA includes elements calculated at netting set and portfolio level. This means that the assessment of optimal exercise needs to be done at least at netting set level\footnote{Although the regulatory capital framework includes elements like the leverage ratio that apply at portfolio level, it is possible to allocate capital appropriately to portfolios as demonstrated by \citet{tasche2008capital}. \citet{green2014portfolio} demonstrate that regulatory capital can be allocated to netting set, including the leverage ratio.}. Hence equation (\ref{e:ex2}) should be considered to apply to the netting set and not the individual trade. Historically, XVA has been computed separately as a valuation adjustment to the main portfolio value. In the case of option products optimal exercise is coded into the valuation model on the basis of equation (\ref{e:ex1}) and typically not into the XVA model, which values the option as though there was no impact on exercise from XVA. This is clearly incorrect and problematic for a number of reasons:
\begin{itemize}
\item The total valuation of the portfolio with the trade including XVA will be wrong.
\item The decision to exercise the option may be incorrect if XVA is not considered leading to suboptimal exercise and potential losses.
\item The risk management of the option in the period close to the exercise date will also be wrong as the combined valuation and XVA sensitivities will be wrong
\end{itemize} 

This paper corrects this problem by proposing models which account for the impact of XVA in option exercise decisions through the use of regression-based techniques. The paper clearly demonstrates that XVA is a significant element in the exercise decision and hence valuation of all options where the XVAs of the underlying trade are material. This includes options traded under CSAs and trades which are cleared, where initial margin and hence margin valuation adjustment (MVA) plays a role.  

Examples are presented along with the numerical impact for typical trades and counterparty portfolios to allow an assessment of the magnitude of the impact of XVA effects on the exercise boundary. The chosen examples are:
\begin{enumerate}
\item An unsecured physically settled European swaption.
\item A physically settled European swaption, where the swap is cleared and therefore subject to MVA
\end{enumerate}

\section{Pricing the Option with XVA at the Exercise Boundary}

\subsection{Economic Value in the Presence of XVA}

\citet{burgard2013funding}, as extended by \citet{green2014kva} and \citet{green2015mva}, derive a PDE for the economic value , $\Vhat$, in the presence of counterparty risk, capital and initial margin:
\begin{align}\label{eq:PDE}
0 = & \frac{\partial \Vhat}{\partial t} +\frac{1}{2} \sigma^2 S^2 \frac{\partial^2 \Vhat}{\partial S^2} - (\gamma_S - q_S) S \frac{\partial \Vhat}{\partial S} - (r + \lambda_B + \lambda_C) \Vhat\nonumber\\
& +g_C \lambda_C + g_B \lambda_B - \epsilon_h \lambda_B - s_X X - \gamma_K K + r \phi K + s_{I_B} I_B - r_{I_C} I_C\\
\Vhat (T, S) = & H(S),
\end{align}
where the notation is summarised in table \ref{tab:not} and close-out conditions on the default of the counterparty (C) and the issuer (B) are given by
\begin{align}
\Vhat(t,S,1,0)&=g_C(M_C,X)\\
\Vhat(t,S,0,1)&=g_B(M_B,X)
\end{align}
respectively, where under standard close-out conditions, $g_C$ and $g_B$ are given by, 
\begin{align}
g_C&=R_C(V-X)^+  +(V-X)^- +X\\
g_B&=(V-X)^+ + R_B(V-X)^- +X
\end{align}
\begin{table}
\centering
\begin{tabular}{|l|p{10cm}|}\hline
{\bf Parameter} & {\bf Description}\\\hline
$\Vhat(t, S)$ & The value of the derivative or derivative portfolio (including valuation adjustments)\\
$V$ & The risk-free value of the derivative or derivative portfolio (excluding valuation adjustments)\\
$U$ & The valuation adjustment\\
$X$ & Collateral\\
$S$ & Underlying stock\\
$r$ & risk-free rate\\
$r_C$ & Yield on counterparty bond\\
$r_X$ & Yield on collateral position\\
$\gamma_S$ & Stock dividend yield\\
$q_S$ & Stock repo rate\\
$\lambda_C$ & Effective financing rate of counterparty bond $\lambda_C = r_C - q_C$\\
$\lambda_B$ & Instantaneous probability of bank default (hazard rate)\\
$s_X$ & Spread on collateral, $s_X = r_X - r$\\
$\epsilon_h$ & Hedging error on default of the issuer\\
$H$ & Derivative payoff\\
$K$ & Capital Requirement\\
$\gamma_K(t)$ & The cost of capital (the assets comprising the capital may themselves have a dividend yield and this can be incorporated into $\gamma_K(t)$\\
$\phi$ & Fraction of capital available for derivative funding\\
$I_B$ & Initial margin posted to counterparty (assumed to be held in a segregated account)\\
$I_C$ & Initial margin received from counterparty (assumed to be held in a segregated account). Note that if the counterparty is a CCP then $I_C=0$\\
$r_{I_B}$ & Yield on posted initial margin\\
$r_{I_C}$ & Yield on received initial margin\\
$s_{I_B}$ &  Spread on posted initial margin $s_{I_B}=r_{I_B} - r$\\
$g_B,g_C$ & Close-out conditions on default of B and C\\
$R_B,R_C$ & Recovery rates on B and C\\
$s_F$     & Funding spread $s_F = (1-R_B)lambda_B$\\
$r_B$     & Yield on an issuer bond with recovery rate $R_B$, $r_B = r + s_F$\\\hline
\end{tabular}
\caption{\label{tab:not}Notation used in XVA model.}
\end{table}

In previous work the next step has been to separate the risk-free value $V$ using the Black-Scholes PDE and so derive an expression for the valuation adjustment $U$. However, where XVA must be considered in option exercise decisions it is more appropriate to solve for the economic value directly. Hence we can apply the Feynman-Kac theorem directly to the PDE in equation \eqref{eq:PDE}  to give
\begin{align}\label{eq:XVAdirect}
\Vhat = & \E_t \left[e^{-\int_t^u(r(s) + \lambda_B(s) + \lambda_C(s)) ds} H(S) | \F_t  \right]\nonumber\\
&  - \int_t^T \lambda_C(u) e^{-\int_t^u(r(s) + \lambda_B(s) + \lambda_C(s)) ds} \E_t \left[g_C(u)\right] du\nonumber\\
& -\int_t^T \lambda_B(u)  e^{-\int_t^u (r(s) + \lambda_B(s) + \lambda_C(s)) ds}\E_t \left[g_B(u)\right] du\nonumber\\
& +\int_t^T \lambda_B(u)  e^{-\int_t^u (r(s) + \lambda_B(s) + \lambda_C(s)) ds}\E_t \left[\epsilon_h(u)\right] du\nonumber\\
& +\int_t^T s_X(u) e^{-\int_t^u (r(s) + \lambda_B(s) + \lambda_C(s)) ds} \E_t\left[X(u)\right] du \nonumber\\
& +\int_t^T r_{I_C}(u) e^{-\int_t^u (r(s) + \lambda_B(s) + \lambda_C(s)) ds} \E_t\left[I_C(u)\right] du\nonumber\\
& +\int_t^T e^{-\int_t^u (r(s) + \lambda_B(s) + \lambda_C(s)) ds} \E_t \left[K(u)(\gamma_K (u) - r\phi)\right] du\nonumber\\
& -\int_t^T s_{I_B}(u) e^{-\int_t^u (r(s) + \lambda_B(s) + \lambda_C(s)) ds}\E_t \left[I_B(u)  \right]du.
\end{align}

\subsection{Calculating forward XVAs using Longstaff-Schwartz}

To evaluate the expectation in equation \eqref{eq:XVAdirect} it will be necessary to calculate the economic value at the exercise date of the option as per equation \eqref{e:ex2}. This means that forward XVA values will need to be calculated, but before proceeding to define a mechanism to do this it is useful to define a generic XVA function,
\begin{equation}
\text{XVA}( \alpha, \beta, \gamma, \delta) = -\int_t^T \alpha(u) e^{ -\int_t^u \beta(s) ds} \E_t[\gamma(u)^\delta] du   \label{e:generic}
\end{equation}
For the classic XVAs as defined by \citet{burgard2013funding}, \citet{green2014kva} and \citet{green2015mva} for the funding strategy \emph{semi-replication with no shortfall at default} the parameters are given in table \ref{t:xva}, although this generic form also covers the integrals in equation \eqref{eq:XVAdirect}. 

\begin{table}[tbp]
\centering
\begin{tabular}{l|c|c|c|c|p{2.4cm}}
{\bf XVA}	& $\alpha$ & $\beta$	 & $\gamma$ &	$\delta$ & {\bf Valuation\break Adjustment}\\\hline
CVA & $(1-R_C)\lambda_C$ & $r+\lambda_C+\lambda_B$ & $V(u)$ & $+$ & Credit \\
FVA & $s_F$ & ditto & $V(u)$ & & Funding\\
COLVA${}_X$ & $s_X$ & ditto & $X(u)$ & & Collateral\\
COLVA${}_{I_C}$ & $r_{I_C}$ & ditto & $I_C(u)$ & & Margin\\
KVA & $\gamma_K-r_B\phi$ & ditto & $K(u)$ &	& Capital\\
MVA	& $s_F-s_{I_B}$ & ditto &	$I_B(u)$ & & Margin\\
\end{tabular}
\caption{\label{t:xva}The generic values in equation \eqref{e:generic} for the funding strategy \emph{semi-replication with no shortfall at default} and standard close-out conditions.}
\end{table}

Equation \eqref{e:generic} can be written in discrete form on the time partition $t =
t_0, ..., t_i, ... , t_N = T$ using the Trapezoid rule (or any suitable alternative numeric integration scheme),
\begin{align}
\text{XVA}_j(t) =& -\h \sum_{i=1}^N \left[  \alpha_j(t_{i-1}) e^{ \int_{t_{i-1}}^u \beta_j(s) ds \E_t[\gamma_j(t_{i-1})^\delta] } \right.  \nonumber\\
&\left.  \qquad\qquad {}+ \alpha_j(t_{i}) e^{ \int_{t_{i}}^u \beta_j(s) ds \E_t[\gamma_j(t_{i})^\delta] } \right] \Delta t_i,
\end{align}
where $j$ indexes the individual XVA terms and $\Delta t_i = t_i - t_{i-1}$. Hence we can write
\begin{equation}
U(t) = \sum_{j=1}^M  \sum_{i=1}^N w_{ij} \E[\gamma_j(t_i)^{\delta_j} | \F_t].
\end{equation}
Recall the optimal exercise condition in the presence of XVA, given by equation (\ref{e:ex2}). In is now apparent that to evaluate these XVA values at some future exercise date, we need to evaluate conditional expectations of the form,
\begin{equation}
\E[ \gamma_j(t_i)^{\delta_j} | \F_t ]
\end{equation}
Conditional expectations like this are readily estimated by regression techniques as described in \citet{green2014portfolio} and \citet{kenyon2015efficient}, indeed the generic regression framework described by \citet{kenyon2015efficient} offers considerable advantages in the case. If we use a Monte Carlo simulation with $N$ time steps and an option exercise at index $n$ then we must evaluate,
\begin{align}
U_{ex}(t_n) =& \sum_{j=1}^M  \sum_{i=1}^N w_{ij} \E[\gamma^{ex}_j(t_i)^{\delta_j} | \F_{t_n}]  \label{e:ex} \\
U_{noex}(t_n) =& \sum_{j=1}^M  \sum_{i=1}^N w_{ij} \E[\gamma^{noex}_j(t_i)^{\delta_j} | \F_{t_n}]  \label{e:noex}
\end{align}
that is we must obtain regression approximations for $2\times(N-n)\times M$ conditional expectations. For the generic approach of \citet{kenyon2015efficient} this simply means layering these additional regressions on top of the existing regressions used to identify the basic trade valuations. Note of course that these regressions are now for the $\gamma_j(t_i)$ not the individual trade values.

The adjustment for the modified exercise boundaries that take into consideration the XVA at the time of exercise can be achieved by a simple modification of the backward induction step in the Longstaff-Schwartz regression. In the second forward Monte Carlo simulation, on option exercise dates, the exercise decision is evaluated simply evaluated using equation \eqref{e:ex2} and the regression approximations to equations \eqref{e:ex} and \eqref{e:noex}. Furthermore it is important to emphasise that this regression is applied across all trades in the relevant portfolio.

\subsection{Portfolio complexities}

The focus of this paper is on the change in the optimal exercise boundary due to credit, funding and capital effects.  This is observed even for single European options that exercise into physical swaps and the effect is material as our numerical examples in Section \ref{s:ex} demonstrate.  A full treatment of the portfolio (potentially bank-wide) problem is outside the scope of this paper but we provide an outline here.

At high level the portfolio solution is the same as the solution for a single trade, i.e. use backwards induction.  The key to modelling the portfolio case is the correct identification of the decision state space.  

Consider two European interest rate swaps (IRS) exercising into physical swaps, say a 3Y-into-5Y (3x5) and a 2Y-into-4Y (2x4).  At maximum portfolio maturity (8Y) the 5Y IRS will mature if the 3x5 was exercised.  On the exercise date of the 3x5 the decision will be affected by whether the 2x4 was exercised or not.  Thus the decision state space at 8Y has four values $\{(0,0),\ (0,1),\ (1,0),\ (1,1)\}$ where the first index states whether the 2x4 was exercised and the second whether the 3x5 was exercised.  Thus we see that $n$ European-into-physical options produce a decision state space with $2^n$ states for the backwards-induction algorithm.

Alternatively, consider two Bermudan IRS exercising into physical swaps with annual exercise dates with the first exercise in 1Y, say both 7Y long (so the last exercise date is in 6Y) and exercising into 7Y-xY swaps, where x is the exercise year. At the end of 7Y two physical swaps may have matured.  Each physical swap could be one of six possibilities.  Thus $n$ Bermudan-into-physical options each with $m(i)$ exercise dates produce a decision state space with $\prod_i^n (m(i)+1)$ states for the backwards-induction algorithm.

Efficient handling of these exponentially-sized decision state spaces of the portfolio problem is outside of the scope of the current paper.  It is a potential application area for quantum computing.  However, conceptually nothing has changed --- we are still applying backwards induction.

\section{Numerical Examples\label{s:ex}}
In this section we consider two examples,
\begin{enumerate}
\item An unsecured physically settled European swaption. In this case the XVAs associated with unsecured derivatives, that is CVA, FVA and KVA, will apply to the underlying swap. 
\item A physically settled European swaption, where the swap is cleared and therefore subject to MVA. In this case the presence of initial margin will limit counterparty exposure and hence CVA, FVA and KVA will be small. However, MVA, the cost of funding the initial margin becomes significant. This is an important case as prior to the introduction of initial margin, XVAs on secured trades were often ignored or considered to be of secondary importance.
\end{enumerate}
using the model setup as described in section \ref{sec:setup}. The numerical results are presented in section \ref{sec:results}. 

\subsection{Setup}\label{sec:setup}
\begin{figure}[htbp]
\centering
\includegraphics[width=0.95\textwidth,clip,trim=50 190 50 50]{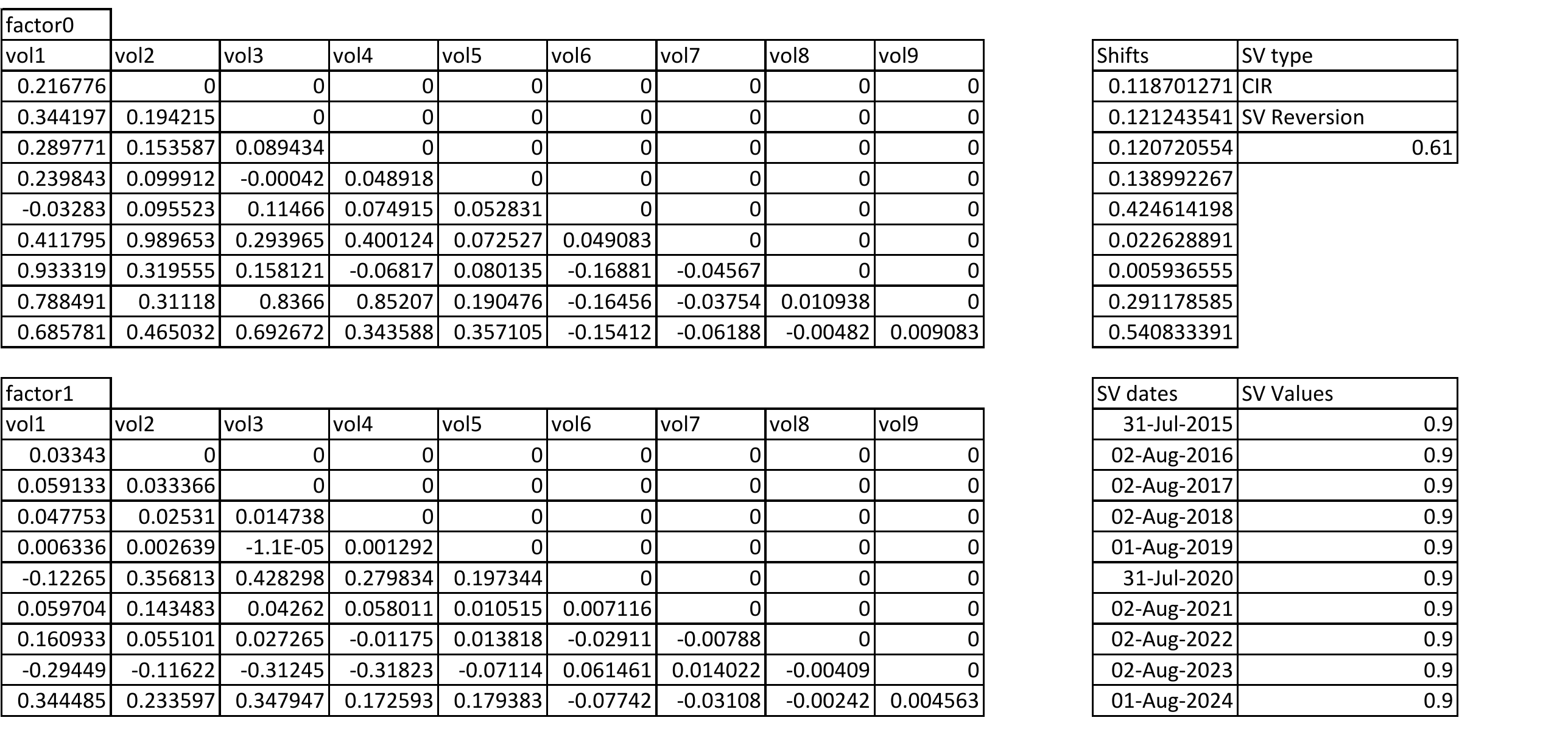}
\caption{	\label{f:param}Shifted 2-factor LMM with flat CIR stochastic volatility, IR model calibration.}
\end{figure}

\begin{figure}[htbp]
	\centering
		\includegraphics[width=0.8\textwidth,clip,trim=50 250 50 250]{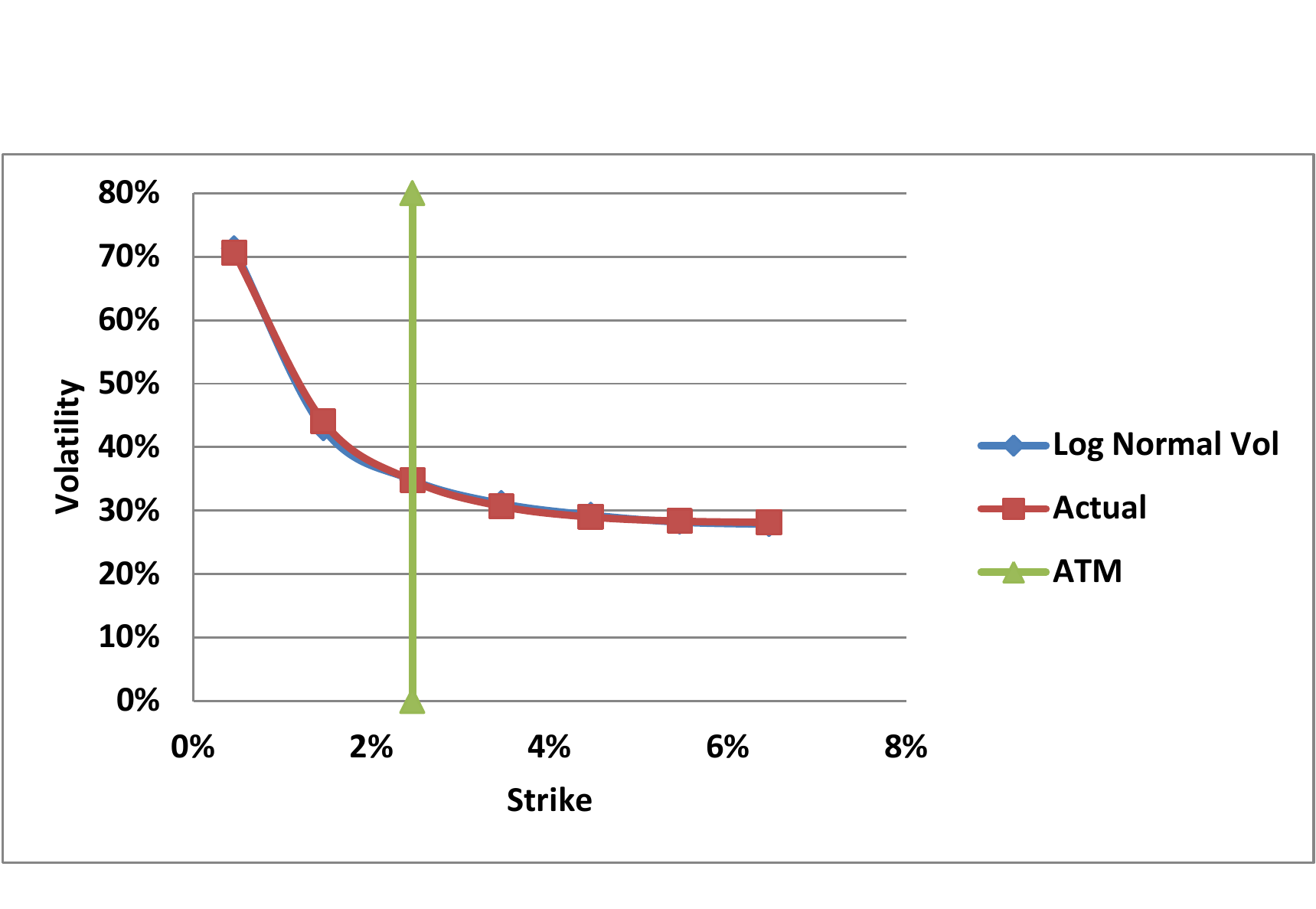}
	\caption{\label{fig:lncalib}Lognormal implied volatility comparison of model versus market. The vertical line gives the at-the-money strike.}
\end{figure}

\begin{figure}[htbp]
	\centering
		\includegraphics[width=0.8\textwidth,clip,trim=50 250 50 250]{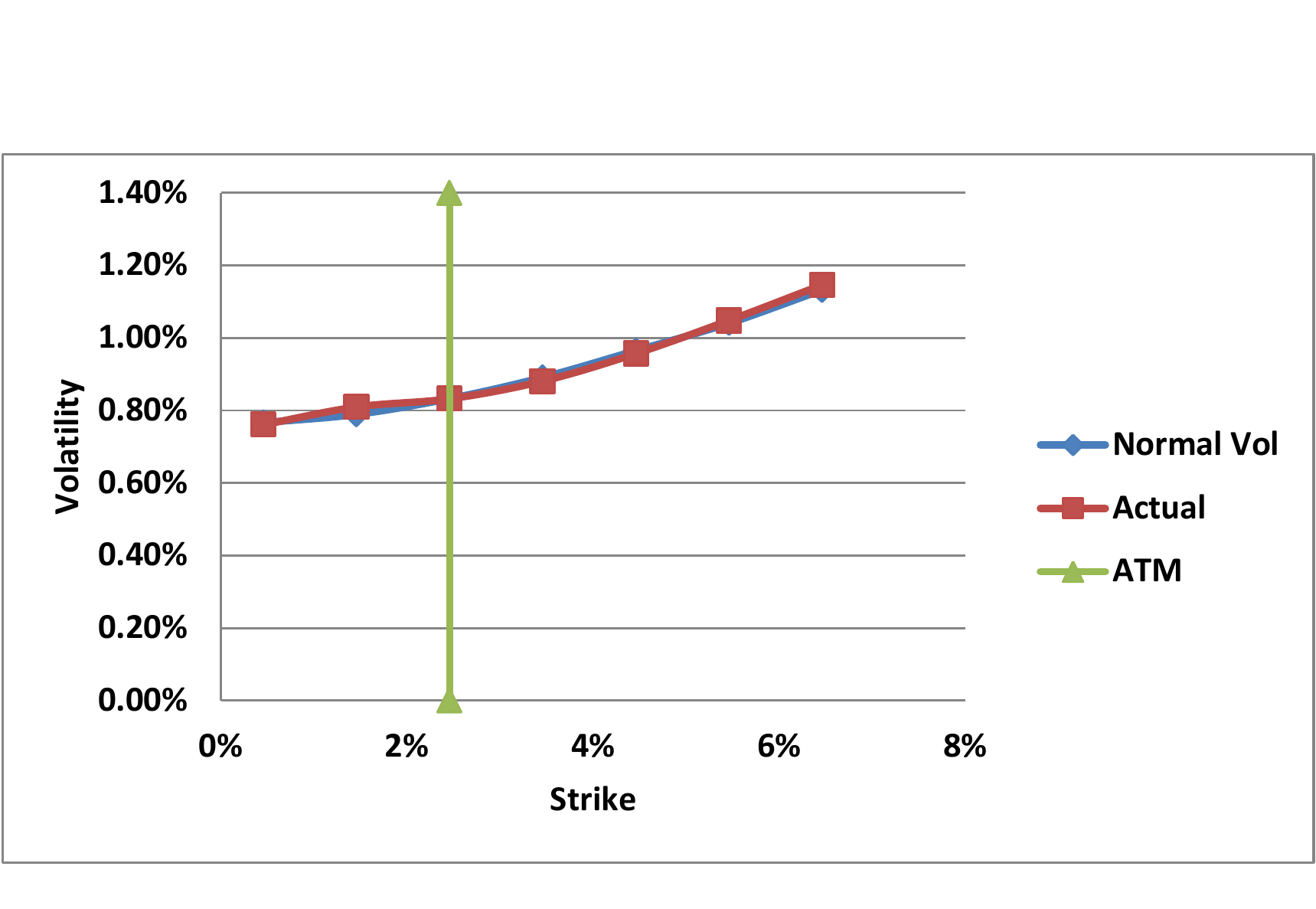}
	\caption{\label{fig:ncalib}Normal implied volatility comparison of model versus market. The vertical line gives the at-the-money strike.}
\end{figure}

The interest rate model that wwas a shifted 2-factor LMM with flat CIR stochastic volatility. The forward rate dynamics are given by 
\begin{equation}
dL_n(t) = ... +(L_n(t) + s_n)\sqrt{x(t)}\lambda_n(t).dW(t)
\end{equation}
where the stochastic volatility is common to all forward rates and follows a CIR process:
\begin{align}
dx(t) =& \theta(1-x(t))dt + \eta(t) \sqrt{x(t)} dZ(t)  \\
dZ dW =& 0\\
x(0) =& 1
\end{align}
This is a standard model available in Numerix CrossAsset\footnote{Details presented with permission from Numerix LLC.  This does not constitute endorsement of Numerix products, nor the reverse.}, see \cite{mihai2013lmm} for details. The model was calibrated to GBP data as of the 31st July 2015 with a dual curve set-up, that is, with SONIA discounting. The smile was calibrated to the 5 year into 5 year smile and to 10 year co-terminal swaptions. The calibrated model parameters are given in figure \ref{f:param}. The fit of the model to the smile is given in terms of lognormal volatilities in figure \ref{fig:lncalib} and in terms of normal volatilities in figure \ref{fig:ncalib}. 

The Monte Carlo simulation used 4096 paths with monthly time steps out to 10 years (121 dates). 

The elements of XVA were calculated using semi-replication with no shortfall as defined in \citet{burgard2013funding} and \citet{green2014kva} and \citet{green2015mva}. The initial margin methodology as applied in the context of MVA used a 97.5\%\ 10-day expected shortfall from 5-day overlapping shocks scaled via $\sqrt{2}$ to 10-day.  Shocks are absolute on tenor point zero yields. A 2500 shock series was used, that is, 10 years of data. The regulatory capital framework assumed was the standardized approach as used in \citet{green2014kva} for a BB-rated counterparty with one modification. Market risk used the \emph{Scenario approach} as defined in \citet{bcbs-128} in section 718(LXiii --- LXV), using a 3-by-3 grid with $\pm25\%$ relative volatility change, and absolute zero yield change from lookup table of equivalent yield moves for interest rates using the table in 781(iv) as specified.  This table depends on the coupon and maturity.

The option price in the presence of XVA was calculated using a two-phase Longstaff-Schwartz regression, with the first phase running from swap maturity to option maturity, and the second phase running from option maturity to the start at $t=0$. Different regressors were used in each phase; in the first phase the local average discount-to-maturity was used while in the second phase the regressor was the discount factor to $t_{\text{now}}$. Local regression used throughout using an averaging filter on all coordinates (bandwidth 15 points). 

To check the accuracy of the regression in the first phase the regression was changed from local to quadratic regression and non-material differences were observed. To assess the accuracy of the second phase we calculated the value of the European options with no valuation adjustments by regression and backwards induction compared to with the expected value which (because the option was European) could be calculated with a single step to maturity. Option price errors were 1\%\ to 3\%\ (relative) across the full range of strikes.  We expect this scaling to hold for valuation adjustments as there was no option exercise decision in the second phase but only backwards induction.  In fact the estimator will be unbiased in this case, that is with no effect from Jensen's inequality and details of this can be found in sections 8.29-8.31 of \cite{glasserman2004montecarlo}.  These errors would reduce with more paths. The simulated 5-into-5 swaption at-the-money strike was 2.44\%\ versus market 2.46\%.

\subsection{Results}\label{sec:results}

\subsubsection{Case 1: Stand alone unsecured physically settled European swaption}

\begin{figure}[htbp]
\centering
\includegraphics[width=0.99\textwidth,clip,trim=0 0 0 0]{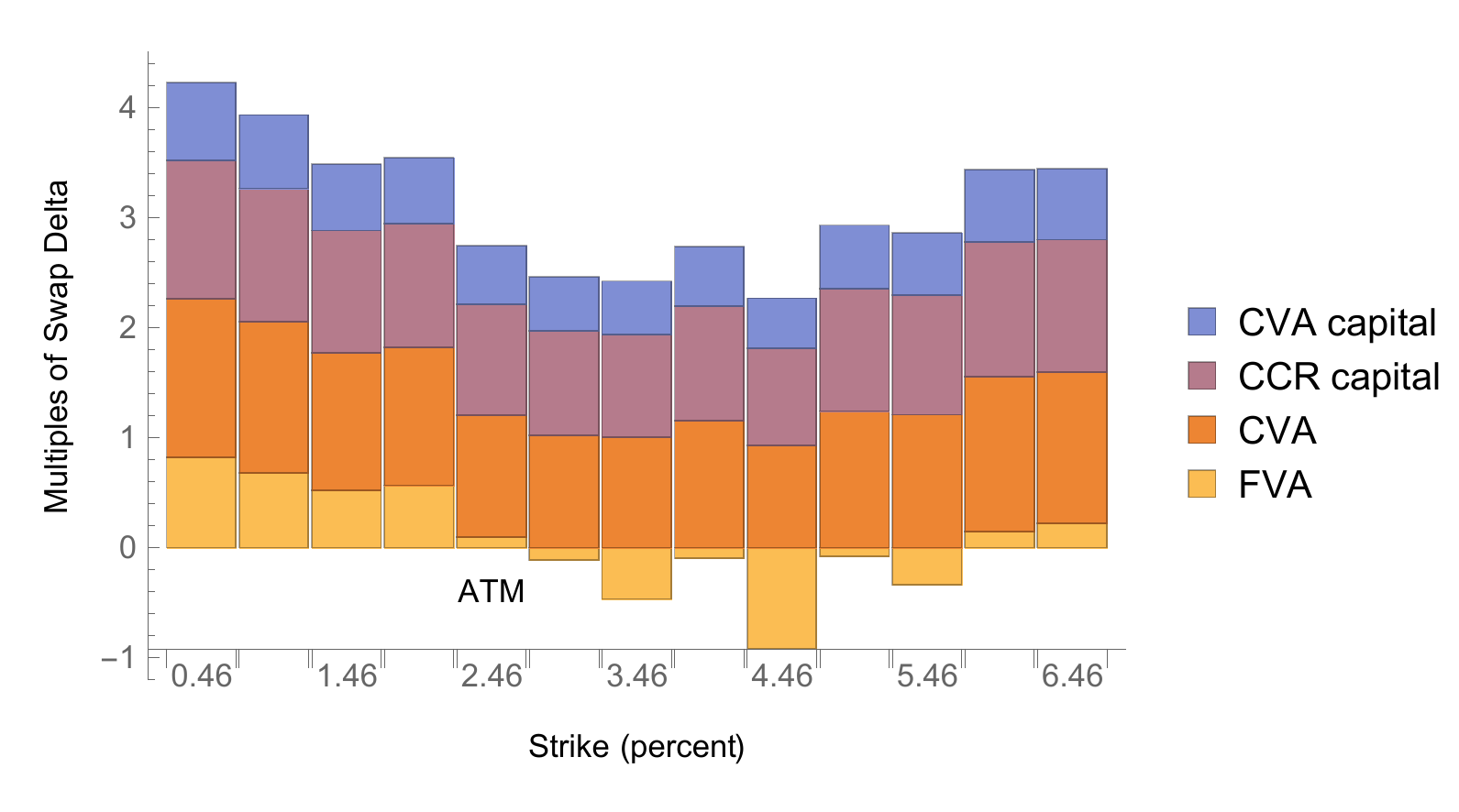}
\caption{\label{fig:55XVAnoMR}XVA of the 5 x 5 European swaption at the time of exercise as a function of strike and swap 01. Market Risk capital is excluded. The horizontal axis is the strike of the swaption in percent with 2.46 being the at-the-money strike. The vertical axis gives the size of XVA expressed as a multiple of the swap delta for the underlying swap and this was approximately 4.5bp.}
\end{figure}

\begin{figure}[htbp]
	\centering
		\includegraphics[width=0.99\textwidth,clip,trim=0 0 0 0]{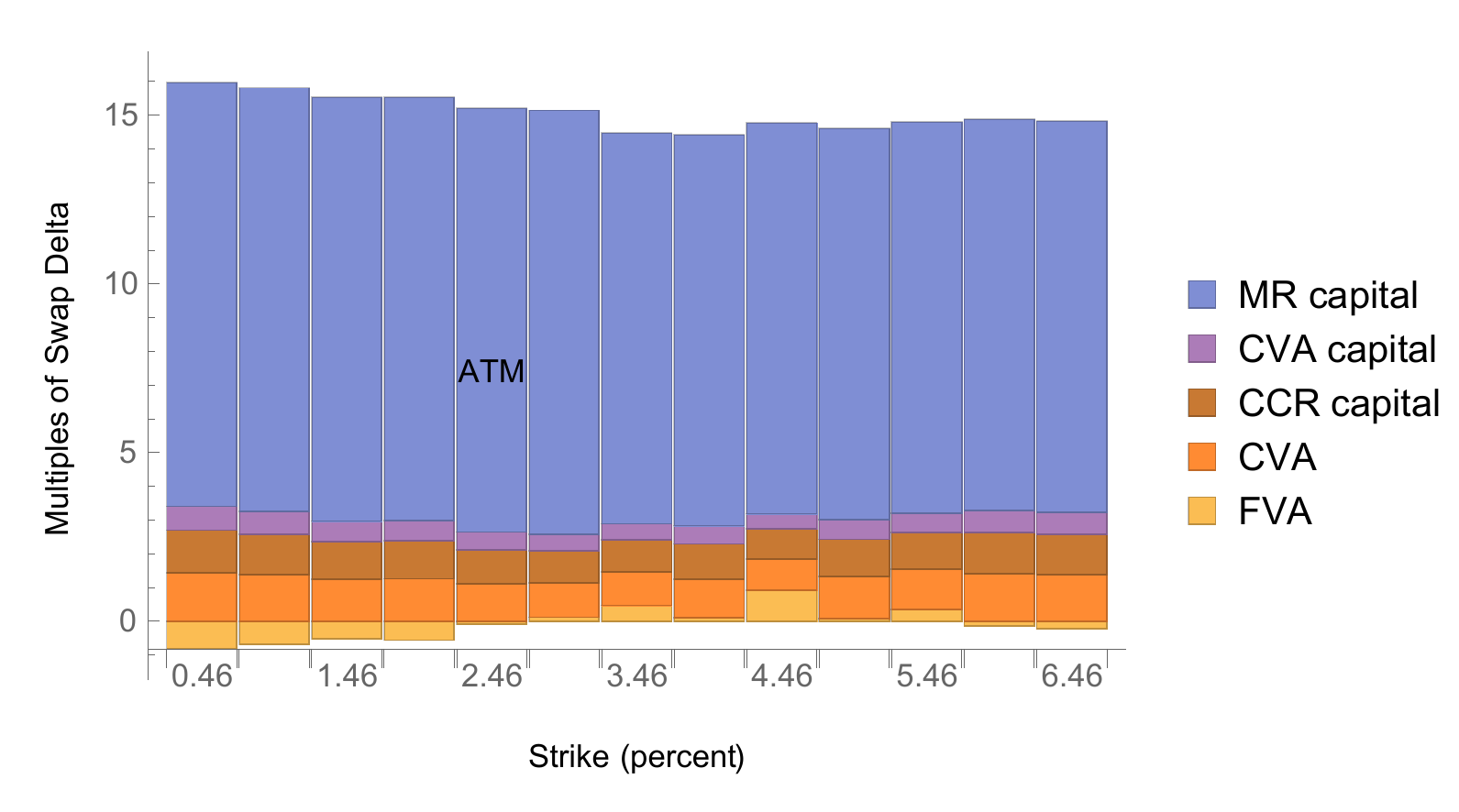}
	\caption{\label{fig:55XVAMR}XVA of the 5 x 5 European swaption at the time of exercise as a function of strike and swap 01. Market Risk capital is included. The horizontal axis is the strike of the swaption in percent with 2.46 being the at-the-money strike. The vertical axis gives the size of XVA expressed as a multiple of the swap delta for the underlying swap and this was approximately 4.5bp.}
\end{figure}

In case one, a stand alone unsecured physically settled European swaption was considered and the size of the XVAs considered at the exercise boundary. These XVAs were calculated for a series of strikes and expressed in terms of the underlying swap delta, which was approximately 4.5 basis points at the exercise date. Figure \ref{fig:55XVAnoMR} shows the size of the component XVAs in terms of the swap delta, excluding market risk. The XVA is approximately 3-4 times the swap delta and hence approximately 13.5-18 basis points in total. Figure \ref{fig:55XVAMR} also includes the impact of market risk capital, which is very large given the trade is standalone and hence unhedged. 

\begin{figure}[htbp]
	\centering
		\includegraphics[width=0.8\textwidth,clip,trim=0 0 0 0]{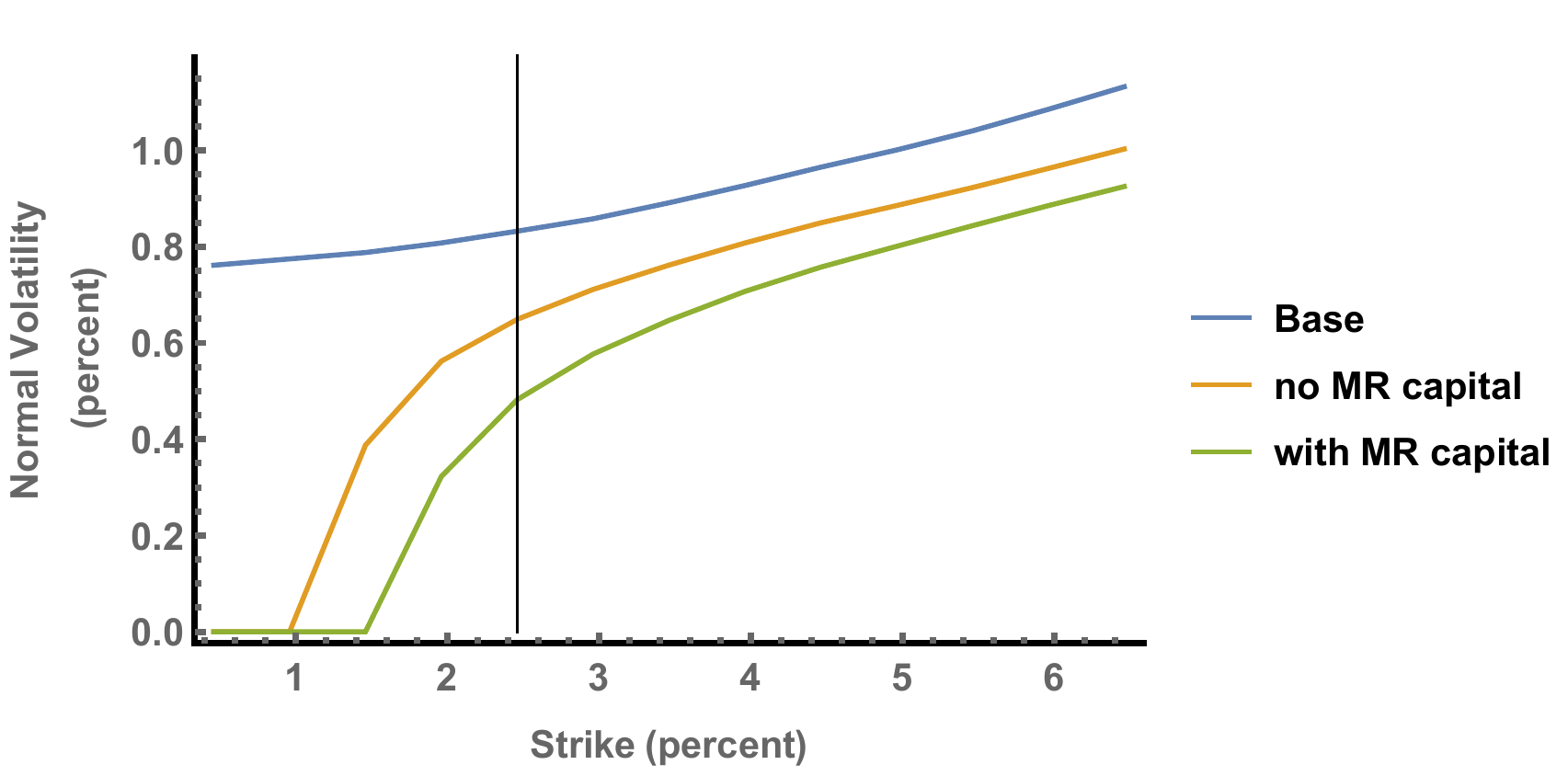}
	\caption{\label{fig:XVASmile1}The vertical line is ATM = $2.46\%$.  Blue line is cash-settled (i.e. no effect of XVA).  The orange line is with XVA except market risk, while the green includes market risk.  In some cases there is no solution, i.e. expect to never exercise}
\end{figure}

The impact of the XVA can also be viewed in terms of a smile shift. Figure \ref{fig:XVASmile1} shows the normal implied volatility of 5-into-5 European swaption. The blue line in the figure represents the cash-settled cases, which is assumed to have no XVAs because of the short period between option exercise and cash settlement. The impact of the XVAs for the stand alone unsecured is illustrated by the orange and green lines, with market risk capital excluded from the orange line and included in the green line. The shift is a very significant when viewed in normal implied volatility terms. Below the at-the-money strike, both orange and green lines fall to zero, reflecting that there are no states of the world where the option is worth exercising, once XVA has been included. 

\subsubsection{Case 2: Physically settled European swaption, where the underlying swap must the cleared}

At the time of writing European swaptions are not subject to mandatory clearing, however, interest rate swaps often are. This means that a physically settled European swaption traded under a bilateral CSA today will resolve to a cleared swap if the option is exercised and the both counterparties are subject to a requirement to clear. The swap will therefore be subject to initial margin requirements and hence MVA should be considered at the time of exercise. Of course a similar situation will eventually prevail generally when the rules for bilateral margin are introduced \citep{bcbs-317}. 

\begin{figure}[htbp]
	\centering
		\includegraphics[width=0.8\textwidth,clip,trim=0 0 0 0]{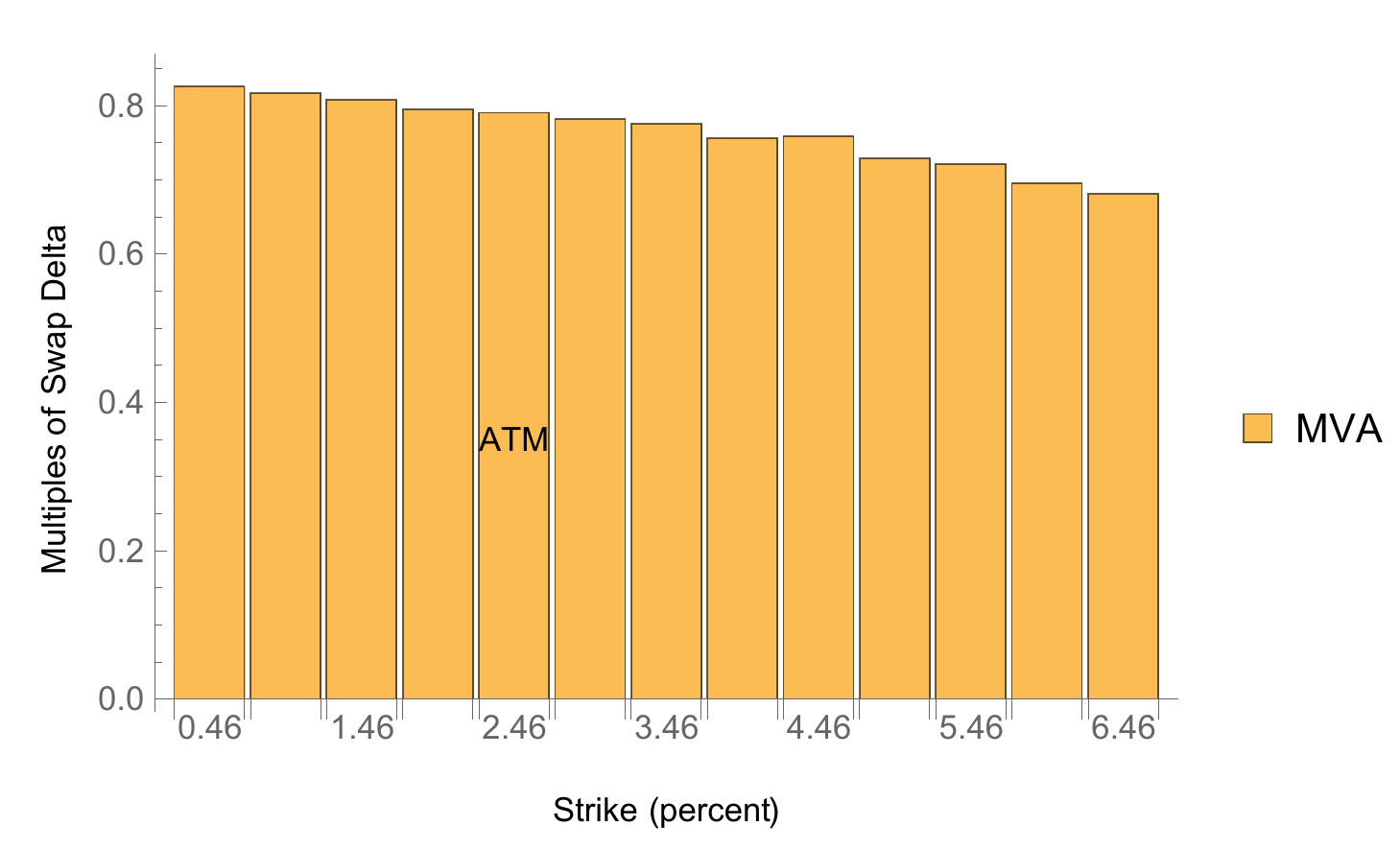}
	\caption{\label{fig:MVAswap01} MVA impact at the time of exercise of the 5 x 5 European swaption, expressed in terms of swap delta. The horizontal axis is the strike of the swaption in percent with 2.46 being the at-the-money strike. The vertical axis gives the size of XVA expressed as a multiple of the swap delta for the underlying swap and this was approximately 4.5bp.}
	\label{p:5}
\end{figure}

\begin{figure}[htbp]
	\centering
		\includegraphics[width=0.8\textwidth,clip,trim=0 0 0 0]{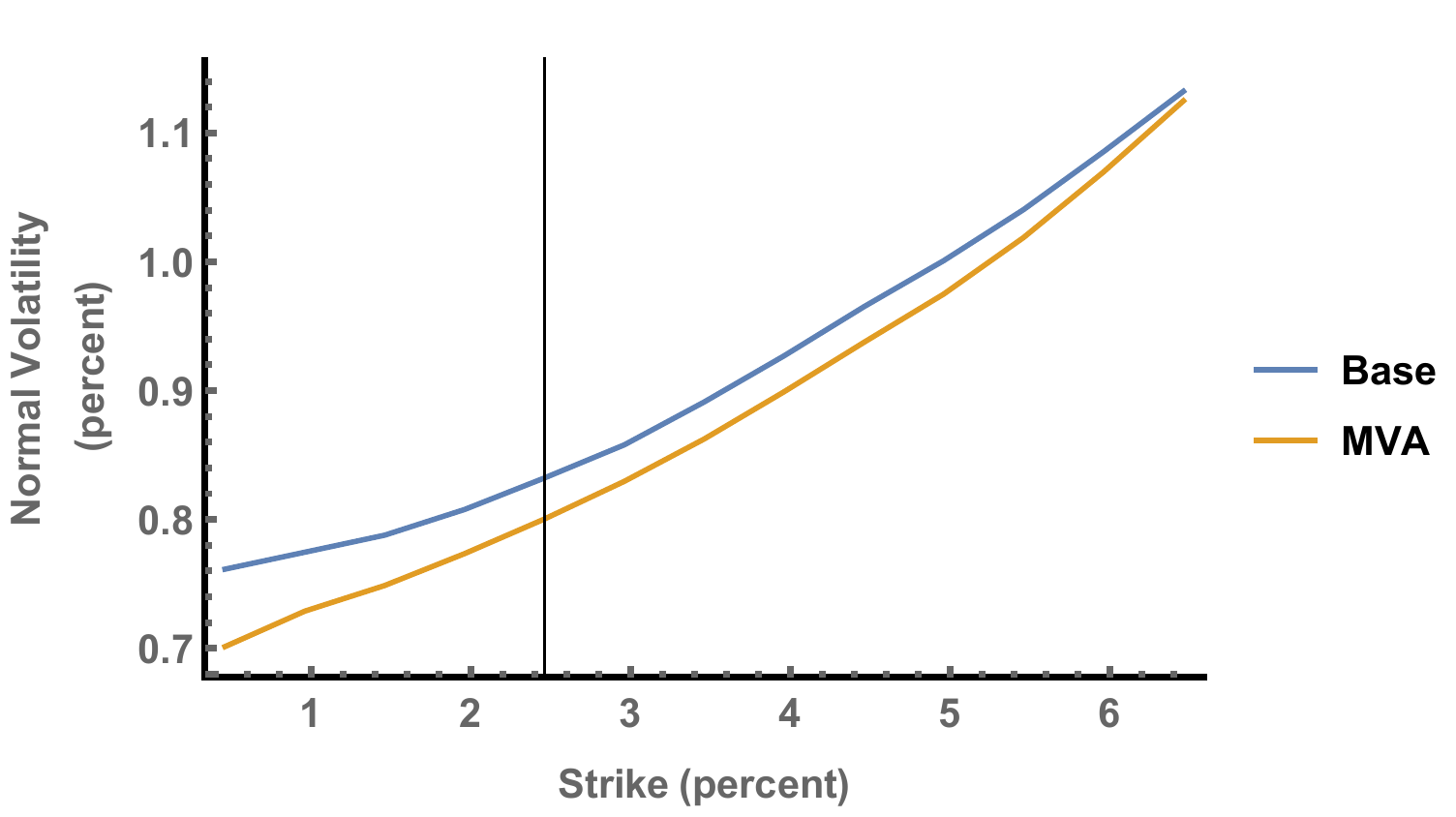}
	\caption{\label{fig:MVAsmile} The vertical line is ATM = $2.46\%$. The blue line is cash-settled (i.e. no effect of XVA). The orange line shows the impact of MVA. }
\end{figure}

In case 2, the MVA was calculated at the time of exercise, based on the initial margin methodology described in section \ref{sec:setup}. Figure \ref{fig:MVAswap01} illustrates the size of the MVA, expressed in terms of the swap delta as a function of the strike. The MVA is between 0.7-0.8 times the swap delta, smaller than the XVAs in the unsecured case but still significant at around 3bp.\footnote{Pre-crisis interest rate swaps would trade with a bid-offer spread of around 0.25 basis points.} Again this can be expressed in terms of a smile shift and this is illustrated in figure \ref{fig:MVAsmile}.   Another view of the effect of MVA is to compare MVA in the option price with the effect of MVA at the point of option exercise.  As shown in Table \ref{t:mva} as the exercise probability drops so does the contribution of MVA to the option price.  Of course the presence of MVA is also changing the probability of exercise.  This illustrates just how important MVA considerations are for proper hedging.

\begin{table}
\centering
\begin{tabular}{|c|c|c|}\hline
Option Strike & MVA at Exercise & MVA in Option Price\\
(percent)&\multicolumn{2}{|c|}{(bps of notional)}\\   \hline
 0.46 & 3.4 & 2.9 \\
 0.96 & 3.5 & 2.7 \\
 1.46 & 3.5 & 2.4 \\
 1.96 & 3.5 & 2.0 \\
 2.46 & 3.5 & 1.8 \\
 2.96 & 3.6 & 1.7 \\
 3.46 & 3.6 & 0.9 \\
 3.96 & 3.7 & 0.6 \\
 4.46 & 3.7 & 0.5 \\
 4.96 & 3.7 & 0.3 \\
 5.46 & 3.8 & 0.2 \\
 5.96 & 3.8 & 0.2 \\
 6.46 & 3.8 & 0.1 \\\hline
\end{tabular}
\caption{Comparison of contribution of MVA to option price and to the option exercise decision.}
\label{t:mva}
\end{table}

\section{Conclusions and Implications}

It is clear from the numerical examples that XVA has a large impact on exercise decisions, and by implication on option valuations and smiles. This is not just an issue for unsecured trades where CVA and FVA have historically been a major consideration, secured and cleared trades are subject to KVA and MVA and hence also to the impact of XVA on optimal exercise. In some cases the impact is so large as to make the option worthless, even though it would have value in the absence of XVA. 

In addition there are clear implications from the netting set and portfolio elements of XVA. Option exercise decisions cannot be taken on the basis of consideration of the option on a stand alone basis, rather at the level of the netting set at a minimum. This can also lead to additional complexity where there are multiple options exercising on the same day. In such circumstances all possible combinations of exercise and non-exercise must be considered to determine optimality which grows exponentially. Except for special cases with non-material XVAs at the time of exercise, the day of stand-alone option pricing models is over. 

\section*{Acknowledgements}

The authors would like to acknowledge feedback from participants at RiskMinds International (Amsterdam, December 2015) and at the WBS fixed income conference (Paris, October 2015) where an earlier version of this work was presented.

\bibliographystyle{chicago}
\bibliography{XVAboundary}

\end{document}